\documentclass[aps,prd,twocolumn,preprintnumbers,amssymb,nobibnotes,nofootinbib,longbibliography,superscriptaddress]{revtex4-2}
\usepackage[utf8]{inputenc}
\usepackage[T1]{fontenc}

\usepackage{amsmath, amsfonts, amssymb}
\usepackage{mathtools}
\usepackage{physics}

\usepackage{graphicx}
\usepackage{svg}

\usepackage{booktabs}
\usepackage{array}

\usepackage{enumitem}

\graphicspath{{figures/}}

\usepackage[%
]{hyperref}

\hypersetup{%
    colorlinks=true,%
    allcolors={blue!60!black}%
}

\usepackage{easy-todo}
\usepackage{cleveref}

\newcommand{\gammacircle}[1]{
  \ensuremath{%
  \gamma_{\circlearrowleft}^{#1}
}}

\newcommand{\dlog}[1]{\dd{\log{#1}}}

\newcommand{\ii}{\,\mathrm{i}\,}
\newcolumntype{C}[1]{>{\centering\arraybackslash}p{#1}}

\begin{document}

\title{Elliptic leading singularities and canonical integrands}

\author{E.~Chaubey}
\affiliation{Bethe Center for Theoretical Physics, Universitaet Bonn, 53115 Bonn, Germany}
\author{V.~Sotnikov}
\affiliation{Physik-Institut, University of Zurich, Winterthurerstrasse 190, 8057 Zurich, Switzerland}

\date{\today}

\begin{abstract}
In the well-studied genus zero case, bases of $\dd\log$ integrands with integer leading singularities define Feynman integrals that automatically satisfy differential equations in canonical form.
Such integrand bases can be constructed without input from the differential equations and without explicit involvement of dimensional regularization parameter $\epsilon$.
We propose a generalization of this construction to genus one geometry arising from the appearance of elliptic curves.
We argue that a particular choice of algebraic one-forms of the second kind that avoids derivatives is crucial.
We observe that the corresponding Feynman integrals satisfy a special form of differential equations that has not been previously reported,
and that their solutions order by order in $\epsilon$ yield pure functions. 
We conjecture that our integrand-level construction universally leads to such differential equations.
\end{abstract}

\maketitle 

\preprint{BONN-TH-2025-18}
\preprint{ZU-TH 29/25}

\section{Introduction}

Feynman integrals are fascinating objects whose properties have broad implications, ranging from precision phenomenology in particle collisions and gravitational wave physics to foundational questions in quantum field theory and cosmology. They also lie at the heart of several open problems where physics and mathematics deeply intertwine.

The method of differential equations (DE) \cite{Kotikov:1990kg,Kotikov:1991pm,Bern:1993kr,Remiddi:1997ny,Gehrmann:1999as}, based upon integration-by-parts identities \cite{Tkachov:1981wb,Chetyrkin:1981qh,Laporta:2000dsw} and the existence of a finite basis of master integrals \cite{Smirnov:2010hn}, has proven remarkably effective in the study of Feynman integrals. In many cases, the DEs can be brought into a so-called canonical form \cite{Henn:2013pwa}, where solutions admit a systematic expansion in the dimensional regularization parameter $\epsilon$, expressed as iterated integrals with logarithmic kernels. This structure enables compact representations of scattering amplitudes \cite{Goncharov:2010jf,Duhr:2012fh,Chicherin:2020oor} and has driven substantial progress in multi-loop calculations. These advances have, in turn, motivated the development of algorithms and tools aimed at systematically identifying the canonical DE form~\cite{Lee:2014ioa,Prausa:2017ltv,Abreu:2018rcw, Meyer:2017joq,Lee:2020zfb,Dlapa:2020cwj}.

It is, however, well established that some Feynman integrals are associated with nontrivial geometry, such as elliptic curves, and cannot be expressed solely through logarithmic differential forms \cite{Adams:2014vja,Sogaard:2014jla}. Such a situation commonly arises in higher-order corrections within the Standard Model involving massive particles, including the top quark, Higgs, electroweak bosons, and even appear in massless theories like maximally supersymmetric Yang-Mills (sYM) \cite{Paulos:2012nu,Caron-Huot:2012awx}. These observations have prompted an extensive search for generalizations of the canonical form~\cite{Caron-Huot:2012awx,Adams:2013nia,Bloch:2013tra,Broedel:2014vla,Bogner:2017vim,Ablinger:2017bjx,Broedel:2017siw,Vergu:2020uur,Bourjaily:2021vyj,Duhr:2024hjf,Duhr:2024bzt,Badger:2024fgb,Driesse:2024feo,Duhr:2024uid,Duhr:2025ppd,Chen:2025hzq,Duhr:2022dxb,Becchetti:2025oyb,Broedel:2018iwv}.

Broadly, two major approaches have emerged in the search for generalized canonical forms.
The first, and so far the most prevalent, begins with an educated guess for the integral basis, derives the associated differential equations, and then seeks transformations that bring them to a desired form, 
typically one that is $\epsilon$-factorized. 
Several mathematically motivated methods have been developed to make this procedure more systematic \cite{brown2017notesmotivicperiods,Adams:2017tga,Gorges:2023zgv,Chen:2025hzq}.

The second approach is viewing Feynman integrals as differential $k$-forms (cohomology) integrated over topologically nontrivial cycles (homology), most naturally interpreted through intersection theory \cite{Mastrolia:2018uzb}.
This perspective, particularly in the context of $\mathcal{N}=4$ sYM theory, has highlighted the special role of $\dd\log$ integrands \cite{Arkani-Hamed:2010zjl,Arkani-Hamed:2010pyv,Arkani-Hamed:2012zlh} and their connection to leading singularities \cite{Cachazo:2008vp,Arkani-Hamed:2010pyv}, which encode a duality between the homology of the punctured Riemann sphere and logarithmic $k$-forms.
The notion of elliptic leading singularities (eLS) was introduced in refs.~\cite{Bourjaily:2020hjv,Bourjaily:2021vyj}, also in the context of sYM theory, offering the first indication of how this geometric framework might extend to elliptic cases.

Elliptic leading singularities are commonly used in modern approaches to guide the construction of the canonical differentials of the first and third kind~\cite{Frellesvig:2021hkr,Gorges:2023zgv,Duhr:2022dxb,Duhr:2025lbz,Becchetti:2025qlu}. While a possibility of similarly constructing the required differentials of the second kind has been mentioned~\cite{Broedel:2018qkq,Gorges:2023zgv,Duhr:2025lbz}, ultimately derivatives are employed for this purpose. Some degree of a posteriori manipulation of the differential equations is also generally necessary.
A systematic approach based exclusively on the integrand-level analysis in fixed integer dimensions remains notably absent.

In this work, we address this gap by revisiting the question: can we construct \emph{integrands} that most naturally reflect the underlying elliptic geometry, with no input from the differential equations that the associated Feynman integrals satisfy?
Pursuing this question leads us to an extension of the method proposed in \cite{Henn:2020lye}, originally developed for identifying $\dd\log$ integrands, to elliptic geometry. 
We begin by choosing a particular basis of algebraic cohomology that is reminiscent to the one employed in the construction of pure elliptic multiple polylogarithms \cite{Broedel:2018qkq}. 
We then match it to a Feynman integral ansatz in a suitable parametrization.
Integration over basis cycles yields the corresponding eLS, which we compute in closed form.
Finally, following the observations in refs.~\cite{Frellesvig:2023iwr,Gorges:2023zgv,Bourjaily:2021vyj},
we perform a simple rotation involving exclusively the eLS that brings the integrands into a form that we define as canonical.
Remarkably, we observe that the associated DEs exhibit a previously unnoticed factorized structure with several intriguing features.
Most notably, the appearance of a mysterious integer whose origin is yet to be understood, and the fact that its solution order-by-order in $\epsilon$ is readily written through iterated integrals over one-forms with (locally) simple poles.
We confirm the existence of such DEs in several state-of-the-art examples and conjecture that our integrand-based construction universally leads to such a form.

\section{Canonical integrands}
Let us define a vector space of Feynman integrals called families by
\begin{equation} \label{eq:loop-momentum-rep}
G ^{\vb{i}} = \int \prod_i \dd^{d}{\ell_i} ~  \frac{1}{\vb*{\rho}^{\vb{i}}},  \qquad  {\vb*{\rho}^{\vb{i}}}= \prod_j \rho_j^{i_j}  
\end{equation}
where $i_j$ are integer indices which can be positive or negative. The $\rho_j$ are Lorentz-invariant polynomials in loop and external momenta, as well as particle masses. 
We denote scalar products of external momenta and masses collectively as $\vb*{x}$. 
The divergences are regularized by deforming the measure to $d = d_0 - 2 \epsilon$ space-time dimensions.
We consider a parametric representation
\begin{equation} \label{eq:parametrization}
  G ^{\vb{i}} = \mathcal{N}(\epsilon) \, \int \dd^{k} \vb*{z} \;\; \mathcal{B}(\vb*{x}, \vb*{z})^{\epsilon} \; g ^{\vb{i}} (\vb*{x}, \vb*{z}),
\end{equation}
with integration over loop momenta exchanged for integration over $k$ loop variables $\vb*{z} = \{z_1, \ldots{}, z_k\}$.
For instance, $\vb*{z}=\vb*{\rho}$ in Baikov representation~\cite{Baikov:1996iu,Frellesvig:2017aai,Bosma:2017ens}.
An exact choice of parametrization, as well as the form of $\mathcal{N}(\epsilon)$ and $\mathcal{B}(\vb*{x}, \vb*{z})$, are not relevant for our discussion.
We call the differential $k$-form $\mathcal{P} \qty(G ^{\vb{i}}) = \dd^{k}{\vb*{z}} ~ \mathit{g} ^{\vb{i}} (\vb*{z})$ the \emph{integrand} of $G ^{\vb{i}}$ in fixed $d_0$ dimensions,
which is the central object of our study.

Given a family of integrals  we can span the space of its integrands by linear combination
\begin{equation} \label{eq:ansatz}
  \sum_{\vb{i}} n_{\vb{i}}(\vb*{x}) ~ G ^{\vb{i}}   ~\xrightarrow{\mathcal{P}}~ \sum_{\vb{i}} n_{\vb{i}}(\vb*{x}) ~ g ^{\vb{i}}(\vb*{x}, \vb*{z}) \,  \dd^k{\vb*{z}}.
\end{equation}
A finite basis of the integrals 
$G ^{\vb{i}}$, known as master integrals (MI), always exists. It is well established \cite{Arkani-Hamed:2010pyv,Cachazo:2008vp,Henn:2013pwa} that analyzing the \emph{integrand}, 
particularly through its leading singularities guides the construction of a good basis.

\subsection{Logarithmic leading singularities}
For completeness, we briefly review the method of Ref.~\cite{Henn:2020lye}. In many relevant cases, one can impose constraints on 
$n_{\vb{i}}$ to bring the integrand ansatz in \cref{eq:ansatz} to the form
\begin{multline} \label{eq:dlog-integrands}
  \sum_{\vb{i}} n_{\vb{i}} ~ g ^{\vb{i}}(\vb*{x}, \vb*{z}) \, \dd^k{\vb*{z}}= \\
  \sum\limits_{j=1}^{n_{\text{LS}}} l_{j} (\vb*{x}, n_{\vb{i}}) \; \dlog{\alpha_{j}^{1} (\vb*{z})} \wedge \cdots \wedge \dlog{\alpha_{j}^{k}(\vb*{z})},
\end{multline}
where the \emph{leading singularities} $l_{j} (\vb*{x}, n_{\vb{i}})$ are linear in the ansatz coefficients $n_{\vb{i}}$,
and $\alpha_{j}^{i}$ are algebraic functions of $\vb*{z}$.
A logarithmic $k$-form on the right hand side represents that the integrand can be fully localized by small circles around zeroes of the algebraic variety $\alpha_{j}^1 = \cdots = \alpha_{j}^k=0$.
Taking the LS corresponds to computing 
\begin{equation} \label{eq:localized-integrand}
  l_{j} (\vb*{x}, n_{\vb{i}})  = \int\limits_{\otimes_i \gammacircle{\alpha_j^i}} \sum_{\vb{i}} n_{\vb{i}}(\vb*{x})~ g ^{\vb{i}}(\vb*{x}, \vb*{z})\, \dd^k{\vb*{z}}.
\end{equation}
This establishes the connection between the space of cycles (homology) and the differential forms (cohomology).
Considering specific cycles that encircle (a subset of) denominators in \cref{eq:loop-momentum-rep} defines \emph{generalized unitarity cuts}.

One then searches for a spanning set $\bar{g}$ of linear combinations \eqref{eq:ansatz} such that all LS are integer,
which amounts to solving a system of linear equations for coefficients $n_{\vb{i}}$,
\begin{equation} \label{eq:LS-equation}
  l_{j} (\vb*{x}, n_{\vb{i}})  = \delta_j^a, \qquad a = \{1,\ldots{},n_{\text{LS}}\},
\end{equation}
which we call \emph{LS equations}.

It has been conjectured \cite{Henn:2013pwa,Herrmann:2019upk} and confirmed in numerous practical applications that the basis of Feynman integrals $\bar{G}$ corresponding to the integrands $\bar{g}$
satisfies a differential equation in the canonical form
\begin{equation} \label{eq:cDE-log}
  \dd{\bar{G}} = \epsilon A ~ \bar{G}, \qquad A = \sum_i A_i \dlog{W_i},
\end{equation}
where $W_i$ are algebraic functions of $\vb*{x}$ and $A_i$ are matrices of rational numbers.

It is well understood, however, that there exist cases where the integrand cannot be fully localized by products of circles as in \cref{eq:localized-integrand}, indicating nontrivial homology,
and presents a geometrical obstruction to the construction outlined above.

\subsection{Elliptic leading singularities}

We now propose a generalization of this method to nontrivial homology.
The simplest obstruction to bringing the ansatz to the form of \cref{eq:dlog-integrands} is when one encounters~\cite{Bourjaily:2017bsb,Broedel:2018qkq,Gorges:2023zgv} 
\begin{equation} \label{eq:elliptic-obstruction}
  \int \frac{ f(z) \dd{z}}{\sqrt{P(z)}} ~\wedge~\dlog{\alpha_1} \wedge \cdots \wedge \dlog{\alpha_{k-1}}, \quad z \equiv z_k
\end{equation}
where $f(z)$ is a rational function and the quadratic polynomial with four distinct roots $r_i$,
\begin{equation}
  P(z) = (z - r_1) (z-r_2) (z- r_3)(z- r_4),
\end{equation}
defines an \emph{elliptic curve}. 
Clearly, the homology in this case cannot be spanned by circles around poles. Instead it is characterized by 
two basis cycles $\gamma_{1}$, $\gamma_2$, which can, without loss of generality, be chosen to enclose the pairs of roots $(r_2,r_3)$ and $(r_3,r_4)$ respectively.
We may also have $m$ cycles $\gammacircle{a_i}$ encircling the simple poles $a_i$ introduced by $f(z)$.

In the following, we assume that the integrand has been already localized by taking $k-1$ residues corresponding to encircling the variety $\alpha_1 = \cdots = \alpha_{k-1}=0$,
and focus on the last integration over $z = z_k$.
The structure of algebraic cohomology in this case is also well understood, and following the conventions of  \cite{Weinzierl:2022eaz}, we define the basis of one-forms as follows:
\begin{equation} \label{eq:integrand-notation}
  \omega_{\kappa}  = \frac{ N_{\kappa}(z) \dd{z}}{\sqrt{P(z)}},
\end{equation}
with
\begin{equation} \label{eq:integrand-basis}
  \begin{aligned}
    &N_\psi  = 1,  \\
    &N_{\phi} =  -\frac{2}{r_{31} r_{42}} \qty(z^2 - \frac{s_1}{2} z + \frac{s_2}{6}  - \frac{1}{6} (r_{21} r_{43}+ r_{31} r_{42})),\\
    &N_{\pi^{\infty}} = z, \\
    &N_{\pi^{a_i}} = \frac{\sqrt{P(a_i)}}{z-a_i} \,.
  \end{aligned}
\end{equation}
We used the notation $r_{ij} = r_i - r_j$ and $s_1 = \sum_{i} r_i$, $s_2 = \sum_{i<j} r_i r_j $.
Here $\omega_\psi$, $\omega_\phi$ are Abel differentials of the first and second kind respectively,
implying that $\omega_\psi$ is holomorphic and $\omega_\phi$  has a double pole at infinity, but no non-vanishing residues.
They correspond to the period and quasi-period of the elliptic curve respectively. Here we constructed the numerator $N_{\phi}$ to integrate to the quasi-period defined in~\cite{Weinzierl:2022eaz}.
$\omega_{\pi^{a_i}}$ is a differential form of the third kind, implying that it has a simple pole at $a_i$ and we normalize it to have unit residue. 

Integration of $\{\omega_\psi$, $\omega_\phi$, $\omega_{\pi^{a_i}}\}$ over the basis cycles ${\gamma_1, \gamma_2}$, yields \emph{elliptic} leading singularities (eLS).
The duality between homology and cohomology can be conveniently visualized by the period matrix
\begin{equation} \label{eq:period-matrix}
  \begin{array}{clccccc}
                 & & \gamma_1 & \gamma_2 & \gammacircle{a_1} & \cdots &  \gammacircle{a_m} \\
    \noalign{\vskip 4pt}
     \hline
    \noalign{\vskip 4pt}
    \omega_\psi & \to \;\;\;\; &  \psi_1 & \psi_2  & 0 &   \cdots & 0 \\
    \omega_\phi &  \to &\phi_1 & \phi_2  & 0 &   \cdots & 0\\
    \noalign{\vskip 5pt}
    \omega_{\pi^{a_1}} & \to & \pi^{a_1}_1 & \pi^{a_1}_2  & 1 &  \cdots & 0\\
    \vdots &  &\vdots & \vdots & \vdots & \ddots & \vdots \\
    \omega_{\pi^{a_m}} & \to  & \pi^{a_m}_1 & \pi^{a_m}_2  & 0 &  \cdots & 1\\
  \end{array}
\end{equation}

The period $\psi_i$ and quasi-periods $\phi_i$ satisfy the coupled DE
\begin{equation} \label{eq:psi-phi-de}
\dd
\begin{pmatrix}
     \psi_i \\
   \phi_i \\
\end{pmatrix}
= 
\frac{1}{2}
\begin{pmatrix}
    \dd\log \frac{1}{r_{32} r_{41}} &
    \dd\log\frac{k}{k-1} \\
    \dd\log \frac{1}{k}  &
    \dd\log r_{31} r_{42} 
\end{pmatrix}
\begin{pmatrix}
     \psi_i \\
   \phi_i \\
\end{pmatrix},
\end{equation}
with $k = \frac{r_{32} \, r_{41}}{r_{31} \, r_{42}}$,
while the differential of $\pi^{a_i}$ is given by
\begin{equation} \label{eq:pi-de}
  \dd \pi_i^{a_i} = \omega^{a_i}_{A}(r_j) ~ \psi_i ~+~ \omega^{a_i}_{B}(r_j) ~\phi_i,
\end{equation}
where the coefficients of $\psi_i$ and $\phi_i$ are rational differential one-forms depending on the roots.
We also have the Legendre identity
\begin{equation} \label{eq:wronskian}
  \psi_1 \phi_2 - \psi_2 \phi_1 = (2 \pi \ii) \, W, \quad W = \frac{1}{r_{31} \, r_{42}}.
\end{equation}
The precise form of the roots and an explicit representation of eLS through complete elliptic integrals of three kinds is shown in \cref{sec:complete-eli}.
Note that sometimes (a linear combination of)  $\omega_\pi^{a_i}$ collapses into a $\dd\log$ form,
in which case its integration over $\gamma_1$ and $\gamma_2$ vanishes, decoupling it from the elliptic curve.
In this case we prefer such linear combination as the basis integrand.

With the cohomology basis in hand, we now address \cref{eq:elliptic-obstruction} as in the logarithmic case.
 Instead of \cref{eq:dlog-integrands}, we bring the ansatz to the form:
\begin{multline} \label{eq:elliptic-integrands}
  \sum_{\vb{i}} n_{\vb{i}} ~ g ^{\vb{i}} \, \dd^k{\vb*{z}}=
  \sum_{j} l_{j} (\vb*{x}, n_{\vb{i}}) \; \omega_{\kappa} \; \bigwedge\limits_{i=1}^{k-1} \dlog{\alpha_{j}^i},
\end{multline}
where $\omega_\kappa$ is one of the basis integrands in \cref{eq:integrand-basis} and the sum is over all different logarithmic $(k-1)$-forms.
We then similarly write the LS equations \eqref{eq:LS-equation} to find a spanning set of linear combinations $\bar{g}_{\mathcal{A}}$ that matches the integrand basis. 
By construction $\bar{g}_{\mathcal{A}}$ have eLS corresponding to \cref{eq:period-matrix}.

Finally, motivated by the observations in refs.~\cite{Frellesvig:2023iwr,Gorges:2023zgv,Bourjaily:2021vyj}, we choose (without loss of generality) a preferred cycle $\gamma_1$ and define the canonical integrands $\bar\omega_\kappa$ as 
\begin{equation} \label{eq:to-canonical}
  \begin{pmatrix*}
  \vphantom{\frac{1}{\psi_1}}\bar\omega_{\psi} \\
  \vphantom{\frac{1}{W} \left(\psi_1\,\omega_{\phi}-\phi_1\,\omega_{\psi}\right)}\bar\omega_{\phi} \\
  \noalign{\vskip 5pt}
  \bar\omega_{\pi^{a_1}} \\
  \vdots \\
  \bar\omega_{\pi^{a_m}} \\
  \end{pmatrix*}
  =
  \begin{pmatrix*}
  \frac{1}{\psi_1} \omega_{\psi} \\
  \frac{1}{W} \left(\psi_1\,\omega_{\phi}-\phi_1\,\omega_{\psi}\right) \\
    \noalign{\vskip 5pt}
  \omega_{\pi^{a_1}} - \frac{\pi^{a_1}_1}{\psi_1} \omega_{\psi}   \\
  \vdots \\
  \omega_{\pi^{a_m}} - \frac{\pi^{a_m}_1}{\psi_1} \omega_{\psi}
  \end{pmatrix*}.
\end{equation}
One can immediately see that the period matrix for $\bar\omega_\kappa$ has the form
\begin{equation} \label{eq:unit-ls-matrix}
  \begin{pmatrix*}
    \mathbf{1} & \frac{\psi_2}{\psi_1}  & 0 &   \cdots & 0 \\
    \mathbf{0} & 1  & 0 &   \cdots & 0\\
    \noalign{\vskip 5pt}
    \mathbf{0} & \;\; \pi^{a_1}_2 -\pi^{a_1}_1 \frac{\psi_2}{\psi_1} \;\;\; & 1 &  \cdots & 0\\
    \vdots & \vdots & \vdots & \ddots & \vdots \\
    \mathbf{0} & \;\; \pi^{a_m}_2 -\pi^{a_m}_1 \frac{\psi_2}{\psi_1}  \;\;\;  & 0 &  \cdots & 1\\
  \end{pmatrix*}
\end{equation}
with all eLS under $\gamma_1$ vanishing, apart from the one corresponding to $\bar{\omega}_\psi$ which is unit.
We suggest that this is a generalization of integrands with unit LS to elliptic geometry.

To summarize, our integrand construction is performed in three steps:
\begin{enumerate}[label=s\arabic*]
  \item\label{step1} Find an integrand parametrization \eqref{eq:parametrization} and bring the ansatz \eqref{eq:ansatz} to the form of \cref{eq:elliptic-obstruction}.
  \item\label{step2} Match it to the basis of algebraic one-forms $\omega_\kappa$ complemented with as many $\dd{\log}$ one-forms as possible
    and solve for ansatz coefficients. This defines \emph{algebraic} basis of integrals $\bar{G}_{\mathcal{A}}$.
  \item\label{step3} Apply the transformation \cref{eq:to-canonical} containing eLS. This defines our final integral basis $\bar{G}$.
\end{enumerate}

We emphasize that employing cuts, while practically useful to simplify \ref{step1}, is not conceptually important.
Indeed, as long as \ref{step1} is accomplished, \ref{step2} only requires solving a system of linear equations. 
Since \ref{step3} concerns only $\omega_\psi, \omega_\phi, \omega_{\pi^{a_i}}$, it should be performed after the basis $\bar{G}_{\mathcal{A}}$ is constructed that is valid without any cuts.

\section{Differential equations}

To study the properties of our integrands in dimensional regularisation, we now calculate the derivatives of the integral bases $\bar{G}_\mathcal{A}$ and $\bar{G}$, and through IBP-reduction~\cite{Lee:2012cn,Peraro:2019svx} derive their DEs.
We assume that all integrals which do not couple to the elliptic curve have already been chosen to have unit LS.

We observe that the algebraic basis $\bar{G}_{\mathcal{A}}$ satisfies a minimally coupled DE linear in $\epsilon$ 
\begin{equation} \label{eq:cDE-alg}
  \dd{\bar{G}_{\mathcal{A}}} = \qty(\tilde{A}_0 + \epsilon \tilde{A}_1) \bar{G}_{\mathcal{A}},
\end{equation}
where $\tilde{A}_0$ can have nonzero entries only in the blocks corresponding to the cuts that reveal the elliptic curve.
These entries match exactly the corresponding logarithmic DE in \cref{eq:psi-phi-de,eq:pi-de} satisfied by eLS.
The latter implies that $A_0$ contains at most $2\times2$ coupled blocks and
$A_1$ has the usual block-triangular structure corresponding to the hierarchy of generalized cuts.

The transformation \eqref{eq:to-canonical} brings \cref{eq:cDE-alg} into the special form where $\epsilon$ 
dependence is factorized component-wise:
\begin{align} \label{eq:cDE}
  \dd \bar{G} =& \; A ~ \bar{G},\quad   A = \; (1 - n\,\epsilon) A_0 + \epsilon  A_1 \nonumber \\  & (A_0)_{ij} \cdot (A_1)_{ij} = 0, \quad A_0^2=0
\end{align}
where no summation over $i,j$ is assumed.
The matrix $A_0$ is strictly upper (or lower) triangular.
We may therefore immediately write $\bar{G}$ order-by-order in $\epsilon$ through Chen's iterated integrals~\cite{Chen:1977oja}.%
\footnote{Note that $A_0=0$ is not necessary for obtaining solutions through iterated integrals \cite{Adams:2018kez}.}
The meaning of integer $n$ remains a mystery to us, and it would be very interesting to understand its origin.
Furthermore, \cref{eq:cDE} satisfies the following properties.
The entries of $A_0$ are precisely the exact one-forms $\dd\qty(\frac{\psi_2}{\psi_1})$, $\dd\qty(\pi^{a_i}_2 -\pi^{a_i}_1\frac{\psi_2}{\psi_1})$ corresponding to the differentials of the period matrix entries for cycle  $\gamma_2$  in \cref{eq:unit-ls-matrix}.
Each entry of $A$ scales homogeneously as $\lambda^k$ under the simultaneous rescaling~\cite{Adams:2018kez} of all eLS, 
$\{\psi_1, \phi_1, \pi^{a_i}_1\} \to \lambda \{\psi_1, \phi_1, \pi^{a_i}_1\}$, with $k \in \{-2, -1, 0, 1, 2\}$. 
We expect this property to be useful for the classification of one-forms, which we leave for future work.
Finally, all one-forms in $A$ have at most simple poles locally around each DE singularity (see \cref{sec:simple-poles} for details).
The solutions therefore corresponds to pure functions as defined in refs.~\cite{Broedel:2018qkq,Frellesvig:2023iwr}.
Let us remark that the DE \eqref{eq:cDE} is explicitly covariant under the change of the preferred basis cycle%
\footnote{The DE for the second cycle is obtained simply by switching the cycle index in all eLS.},
and in practice one must always locally choose the cycle on which the eLS can be expanded into power series around the closest singular point \cite{Frellesvig:2023iwr,Duhr:2025lbz}.
Another useful feature of these functions is that one can easily construct their linear combinations that are invariant under the choice of cycle by applying the inverse of \cref{eq:to-canonical}.

We conjecture that all these properties hold for the bases $\bar{G}_\mathcal{A}$ and $\bar{G}$ constructed by following the method presented in the previous section.

\subsection{On the role of $\epsilon$ factorization}

Many appealing properties of the $\epsilon$-factorized DE in $\dd{\log}$ form \cref{eq:cDE-log} also hold for \cref{eq:cDE}. An evident difference is the identification of the $\epsilon$-expansion order with the maximal length of iterated integrations at that order, corresponding to the algebra's grading by transcendental weight~\cite{Henn:2013pwa}.

We could convert \eqref{eq:cDE} into an $\epsilon$-factorized form by fully diagonalizing the period matrix \eqref{eq:unit-ls-matrix},%
\footnote{Equivalent to ``integrating out'' $A_0$ in \cref{eq:cDE},}
enforcing alignment between $\epsilon$ order and iterated integration length. However, this breaks covariance of \cref{eq:cDE} under preferred cycle changes and rescaling homogeneity. There are indeed indications that one should not expect such alignment with non-logarithmic one-forms \cite{Broedel:2018qkq}. $\epsilon$-factorized DE have also been linked to function classes with appealing properties~\cite{Adams:2017ejb} and may aid in establishing linear dependencies in the solution space~\cite{Deneufchatel:2011yph,Duhr:2024xsy}, crucial for modern analytic amplitude methods~\cite{Chicherin:2020oor,Abreu:2023rco,Badger:2024dxo,Gehrmann:2024tds,DeLaurentis:2025dxw}.

Understanding how these insights relate to the integrand basis proposed here is an interesting direction for future work, alongside the study of the function space of the solutions of \cref{eq:cDE}.

\section{Examples}
\label{sec:examples}

A straightforward application of the method outlined above allowed us to derive the DE in the form of \cref{eq:cDE} for a number of examples that has been recently studied in the literature.
We summarize the results in \cref{tab:examples-summary}.

\begin{table}[ht]
  \newcommand{\picwidth}{10ex}
  \newcommand{\centeredfig}[1]{$\vcenter{\hbox{\includegraphics[width=\picwidth]{#1}}}$}
  \renewcommand{\arraystretch}{2.1}
  \centering
  \begin{tabular}{C{16ex}  C{16ex}   C{10ex}C{10ex}cc}
    \toprule
    Graph & References & Number of MI   & Number of $\pi^{a_i}$ & $n$ \\
    \midrule
    \centeredfig{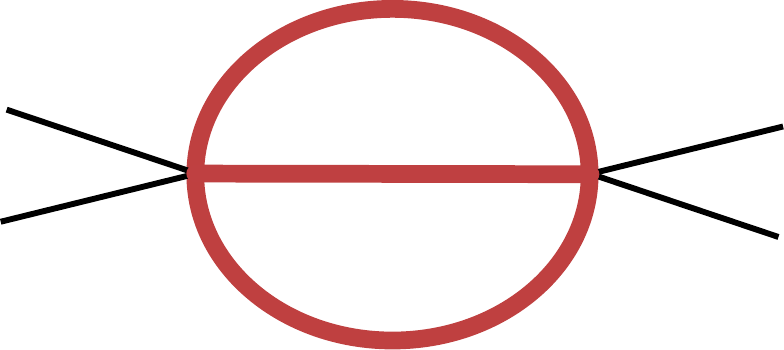}         & \cite{Tarasov:2006nk,Caffo:1998du,Adams:2015pya}  & 2    & 0 & 3   \\
    \centeredfig{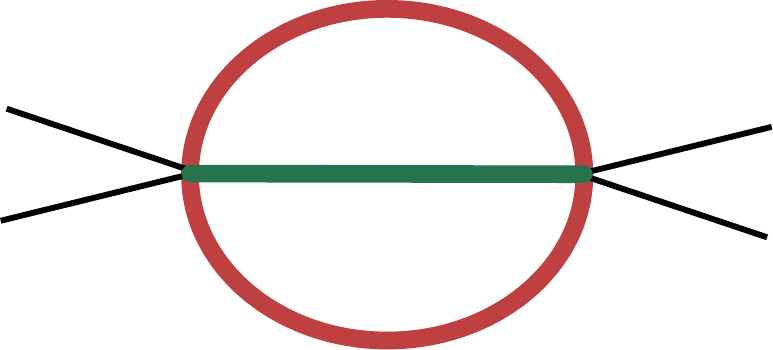}          & \cite{Duhr:2025lbz}  & 3    & 1 & 3   \\
    \centeredfig{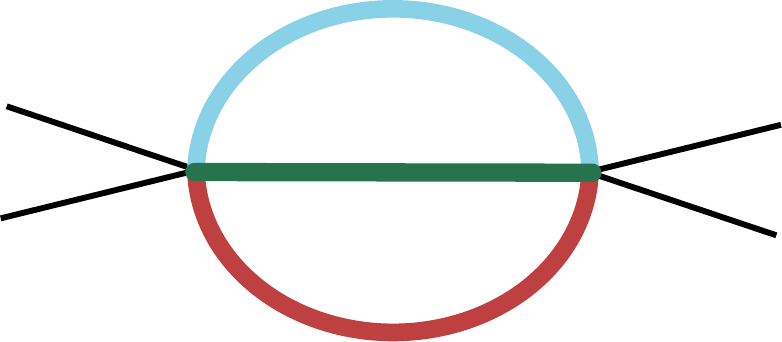}        & \cite{Muller-Stach:2011qkg,Adams:2013nia,Bogner:2019lfa}  & 4    & 2 & 3     \\[1ex]
    $\vcenter{\hbox{ $\begin{multlined}  \includegraphics[height=5ex]{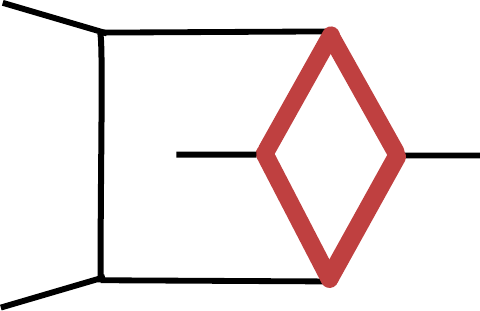} \\[-3ex] \;\includegraphics[height=4.5ex]{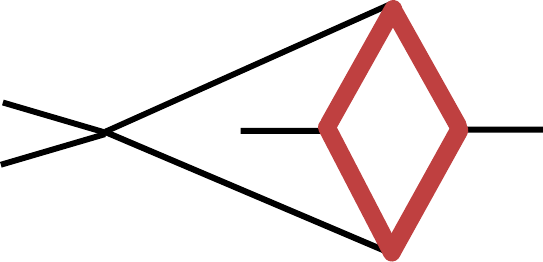} \end{multlined}$  }}$ & \cite{vonManteuffel:2017hms,Becchetti:2023wev,Ahmed:2024tsg,Becchetti:2025rrz,Ahmed:2025osb}   & 4+2  & 1+0 & 4   \\
    \centeredfig{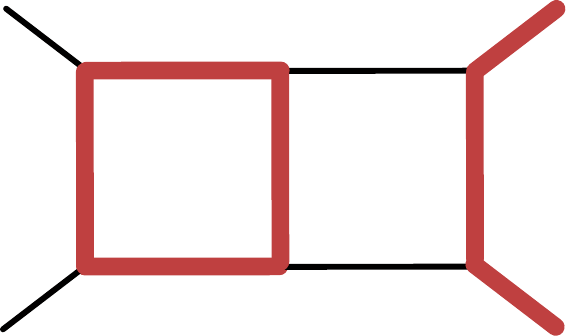}                    & \cite{Adams:2017tga,Adams:2018bsn,Adams:2018kez}  & 5    & 1 & 2   \\
    \centeredfig{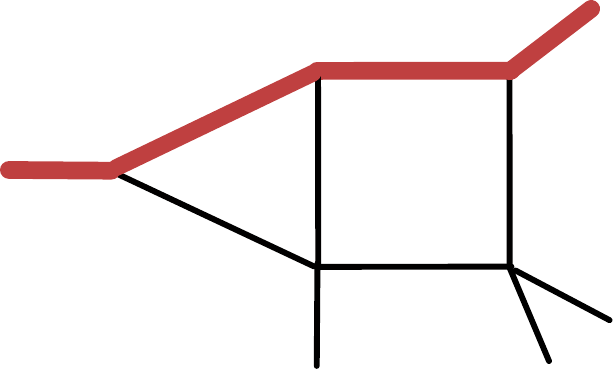}                       & \cite{Badger:2024fgb,Badger:2024dxo,Becchetti:2025oyb}  & 3    & 1 & 2   \\
    \bottomrule
  \end{tabular}
  \caption{Example of Feynman graphs considered in this work. Thick lines denote massive lines (red, blue, green correspond to different masses), $n$ is the integer in $1 - n\,\epsilon$.}
  \label{tab:examples-summary}
\end{table}

For all sunrise diagrams (first three rows) we do not apply cuts, while for the others we consider minimal cuts supporting the corresponding elliptic curve. The number of MI on these cuts, shown in column 3, equals $2$ plus the number of non-vanishing third-kind eLS $\pi^{a_i}$, plus remaining integrals chosen to have $\dd{\log}$ integrands with unit LS.

For example, we define the two unequal mass sunrise graph, in column 2 in \cref{tab:examples-summary}, with $\rho_i= \{(\ell_1)^2-m^2,(\ell_2)^2-M^2,(\ell_1-\ell_2-p)^2-m^2,\ell_1 \cdot p ,\ell_2 \cdot p\}$, where $p$ is the external momentum. In $d_0=2$, we obtain
\begin{equation}
  \bar{G}_{\mathcal{A}} = \begin{pmatrix}
      G^{1,1,1,0,0}  \\
      \sum_{j=0}^{2} c_j \, G^{1,1,1,0,-j}+  4 W \, G^{1,0,1,0,0}  \\
      G^{1, 1, 1, 0, -1} \\
      G^{0, 1, 1, 0, 0} \\
      G^{1, 0, 1, 0, 0} \\
  \end{pmatrix},
\end{equation}
with $c_0 = \frac{1}{2}$, $c_1= 4 s_1 W$, and $c_2 = - 8 W$.
The basis $\bar{G}$ is given by
\begin{equation}
  \renewcommand{\arraystretch}{1.3}
  \bar{G} =
    \begin{pmatrix}
      \frac{1}{\psi} &  0 & 0 & 0 & 0 \\
      -\frac{\phi}{W} &  \frac{\psi}{W} & 0 & 0 & 0 \\
      -\frac{\pi^{\infty}}{\psi} &  0 & 1 & 0 & 0 \\
      0 &  0 & 0 & 1 & 0 \\
      0 &  0 & 0 & 0 & 1 \\
    \end{pmatrix} \; \bar{G}_\mathcal{A},
\end{equation}
where the cycle index is implicit. 
It satisfies the DE \mbox{$\dd\bar{G} = \qty((1 - 3\,\epsilon) A_0 + \epsilon A_1)\,\bar{G}$}, with 
\begin{equation}
  A_0 = \begin{pmatrix}
    0 &  \dd\qty(\frac{\psi_2}{\psi_1}) & 0 & 0 & 0 \\
      0 &  0 & 0 & 0 & 0  \\
       0 &   \dd(\pi^{\infty}_2 -\pi^{\infty}_1\frac{\psi_2}{\psi_1}) & 0 & 0 & 0 \\
       0 &  0 & 0 & 0 & 0 \\
      0 &  0 & 0 & 0 & 0 \\
    \end{pmatrix}\,,
\end{equation}
and $A_1$ schematically given by
\begin{equation}
  \begin{pmatrix}
     \dfrac{p^{(1)}_{\psi, \phi, \pi^\infty}}{\psi}  &  0 & \dfrac{p^{(0)}}{\psi}  & \dfrac{p^{(0)}}{\psi} &  \dfrac{p^{(0)}}{\psi} \\
       p^{(2)}_{\psi, \phi, \pi^\infty}  &  \dfrac{p^{(1)}_{\psi, \phi}}{\psi}  &  p^{(1)}_{\psi, \phi} &  p^{(1)}_{\psi, \phi} &  p^{(1)}_{\psi, \phi}  \\
       \dfrac{p^{(2)}_{\psi, \phi, \pi^\infty}}{\psi} &  0 & \dfrac{ p^{(1)}_{\psi, \pi^\infty}}{\psi} & \dfrac{ p^{(1)}_{\psi, \pi^\infty}}{\psi} & \dfrac{ p^{(1)}_{\psi, \pi^\infty}}{\psi} \\
       0 &  0 & 0 & -2  \dlog{w_1} & 0 \\
      0 &  0 & 0 & 0 & - 4  \dlog{w_2} \\
    \end{pmatrix} \,,
\end{equation}
where $p^{(k)}_{\kappa}$ are distinct homogeneous polynomials of degree $k$ in the eLS indicated as subscripts and algebraic one-forms as coefficients.

We provide the bases $\bar{G}$ for all considered examples in supplementary material.
A more detailed discussion of examples and comparison to the existing methods will be presented in the followup paper.

\section{Discussion and outlook}

In this letter, we introduced a method to construct a basis of Feynman integrands that naturally captures the underlying elliptic geometry, extending earlier construction based on logarithmic forms. 
The resulting integrals satisfy a new factorized form of DE, which we argued possesses several appealing structural properties. 
We believe this offers a new perspective onto the generalization of canonical DE to genus-one geometry and provides a systematic starting point for further investigations.
In particular, we are hopeful that it will offer insights into the construction of special functions bases that reveal the simplicity of scattering amplitudes~--- an aspect that has been essential in the development of modern approaches to multi-scale amplitude calculations in the logarithmic case.

Our construction resembles that of refs.~\cite{Bourjaily:2020hjv,Bourjaily:2021vyj} in sYM theory. However, only $\omega_\psi$ and $\omega_{\pi^{a_i}}$ appear there, lacking a second-kind differential to complete the cohomology. This is explained by noting that $\omega_\phi$ produce UV-divergent Feynman integrals. Thus, in sYM theory, one expects the amplitudes to conspire such that the coefficient of $\omega_\phi$ vanishes.

Our construction offers practical advantages: it is algorithmic, anticipating implementation similar to~\cite{Henn:2020lye,Chen:2022lzr}, and independent of the number of external scales while explicitly covariant in both the treatment of scales and the choice of basis cycles. This makes it especially well-suited for multi-scale scattering amplitude calculations. Without squared denominators, deriving DEs may also be more efficient. For numerics, the algebraic basis is likely optimal for local generalized series methods~\cite{Moriello:2019yhu,Hidding:2020ytt,Armadillo:2022ugh,Prisco:2025wqs}, and closed-form eLS expressions via complete elliptic integrals benefit one-fold integral approaches~\cite{Caron-Huot:2014lda,Chicherin:2020oor,Chicherin:2021dyp,Abreu:2023rco}.

Finally, we highlight key differences between our approach and related methods~\cite{Frellesvig:2021hkr,Gorges:2023zgv,Pogel:2022yat}. Our construction avoids derivatives in defining the algebraic basis, requires no $\epsilon$-dependent rescaling, and directly yields the DE in form~\cref{eq:cDE} without further manipulation. It also directly provides a basis for an $\epsilon$-factorized DE if desired. In contrast, standard approaches typically 
involve some degree of post-processing. We verified that taking the derivative of $\omega_\psi$ in place of $\omega_\phi$, 
while otherwise following all steps of our construction, does decouple the homogeneous 
part of the DE, but does not lead to \eqref{eq:cDE},
producing higher orders in $\epsilon$ in \cref{eq:cDE-alg,eq:cDE}.  Thus, the choice of the second-kind differential in the algebraic cohomology basis is crucial, with the structure of~\cref{eq:psi-phi-de} likely playing a key role.

Several open questions deserve investigation. First, understanding the integer $n$ in~\cref{eq:cDE} is valuable. Clarifying dependencies among the one-forms in \cref{eq:cDE} and their associated functions is another immediate direction. Finally, having focused on the simplest geometry beyond the Riemann sphere, a natural future step is to generalize to higher-genus curves and higher-dimensional algebraic varieties.

\begin{acknowledgments}

We would like to thank Thomas Gehrmann,  Andreas  von Manteuffel, Tong-Zhi Yang, Simone Zoia for inspiring discussions.
We are grateful to Johannes Henn, Harald Ita, Stefan Weinzierl for comments on the manuscript.
V.S.\ extends gratitude to the Bethe Center for Theoretical Physics at the University of Bonn for hospitality. E.C.\ would like to thank the Physics Institute, University of Zurich for hospitality.
  
V.S.\ has received funding from the European Research Council (ERC)
under the European Union's Horizon 2020 research and innovation
programme grant agreement 101019620 (ERC Advanced Grant TOPUP). The work of E.C.\ is
funded by the ERC grant 101043686 ‘LoCoMotive’. Views and opinions expressed are however those of the author(s) only and
do not necessarily reflect those of the European Union or the European Research Council.
Neither the European Union nor the granting authority can be held responsible for them.

\end{acknowledgments}

\vspace*{3em}


\appendix


\section{Representation of elliptic leading singularities through complete elliptic integrals.}
\label{sec:complete-eli}

The elliptic leading singularities $\psi$,  $\phi$, $\pi ^{a_i}$ can be represented in closed form through complete elliptic integrals of first, second, and third kind respectively.
An explicit form depends on the arrangement of roots (see e.g.~\cite{Broedel:2019hyg}). 
For reference we give here a representation for four real roots ordered as $r_1 < r_2 <r_3<r_4$.

Here we employ the following conventions for the complete elliptic integrals,
\begin{align}
  \mathrm{K}(m) &= \int_0^{\frac{\pi}{2}} \frac{\dd\theta}{\sqrt{1 - m \sin^2\theta}}, \\
  \mathrm{E}(m) &= \int_0^{\frac{\pi}{2}} \sqrt{1 - m \sin^2\theta} \, \dd\theta, \\
  \Pi(n, m) &= \int_0^{\frac{\pi}{2}} \frac{\dd\theta}{(1 - n \sin^2\theta)\sqrt{1 - m \sin^2\theta}},
\end{align}
and denote
\begin{equation}
  k =  \frac{r_{32} \, r_{41}}{r_{31} \, r_{42}}, \qquad \bar{k} = 1 - k =  \frac{r_{21} \, r_{43}}{r_{31} \, r_{42}}.
\end{equation}

\begin{widetext}
  The integration over the first cycle $\gamma_1$ yields
  \begin{align}
    \psi_1 &= \frac{2 \, \mathrm{K}\left( k \right)}{\sqrt{r_{31} \, r_{42}}},\\
    \phi_1 &= \frac{2 \, \mathrm{E}\left( k \right)}{\sqrt{r_{31} \, r_{42}}},\\
    \pi_1^{\infty} &= r_1 \, \psi_1 + \frac{2 \, \Pi\left( \frac{r_{23}}{r_{13}}, k \right) \, r_{21}}{\sqrt{r_{31} \, r_{42}}}, \\
    \pi_1^{a} &= \frac{\sqrt{P(a)} }{r_1 - a} \left( \psi_1  +  \frac{\, r_{21}}{(a - r_2) \sqrt{r_{31} \, r_{42}}}\Pi\left( \frac{(r_1-a) \, r_{23}}{(r_2 - a) \, r_{13}}, k \right) \right) \,.
  \end{align}
  And the integration over the second cycle $\gamma_2$ yields
  \begin{align}
    \psi_2 &= \frac{-2\,\mathrm{i} \, \mathrm{K}(\bar{k})}{\sqrt{r_{31} \, r_{42}}}, \\
    \phi_2 &=  \psi_2  + \frac{2 \, \mathrm{i}\, \mathrm{E}(\bar{k})}{\sqrt{r_{31} \, r_{42}}}, \\
    \pi_2^{\infty} &=  r_2 \, \psi_2 - \frac{2 \, \mathrm{i} \, \Pi\left( \frac{r_{34}}{r_{24}}, \bar{k} \right) \, r_{32}}{\sqrt{r_{31} \, r_{42}}}, \\
    \pi_2^{a} &= \frac{\sqrt{P(a)} }{r_2 - a}  \left(\psi_2
      - \frac{2 \, \mathrm{i} \, \, r_{32}}{(a - r_3) \sqrt{r_{31} \, r_{42}}}\Pi\left( \frac{(r_2-a ) \, r_{34}}{(r_3-a) \, r_{24}}, \bar{k} \right) 
    \right).
    \end{align}

In addition, we show an explicit form of the differential of $\pi^{a_i}_i$ expressed through $\psi_i$ and $\phi_i$,
which is useful in derivation of the differential equations:
\begin{align}
  \notag
  \dd{\pi_i^{\infty}} =& -\frac{1}{2} \left(r_3 \, \dd\log\left( \frac{r_{32}}{r_3} \right) + r_4 \, \dd\log\left( \frac{r_{41}}{r_4} \right)\right) ~ \psi_i \quad + \\
                      &\frac{1}{2} \left(
  -r_2 \, \dd\log\left( \frac{r_{21}}{r_2} \right)
  + r_3 \, \dd\log\left( \frac{r_{32}}{r_3} \right)
  + r_4 \, \dd\log\left( \frac{r_{41}}{r_{43}} \right)
  \right) ~ \phi_i, \\
  \notag
  \dd{\pi_i^{a}} =& \frac{\sqrt{P(a)}}{2} \left(\frac{\dd\log\left( \frac{r_{32}}{r_3 - a} \right)}{(a - r_2)} + \frac{\dd\log\left( \frac{r_{41}}{r_4 - a} \right)}{(a - r_1)}\right)  ~ \psi_i  \quad + \\ 
                 & \frac{\sqrt{P(a)}}{2}\left( \frac{\dd\log\left( \frac{r_{21}}{r_2 - a}  \frac{r_4 - a}{r_{41}} \right)}{(a - r_1)}
  + \frac{\dd\log\left( \frac{a - r_3}{r_{32}} \right)}{(a - r_2)}
  + \frac{\dd\log\left( \frac{r_{34}}{r_4 - a} \right)}{(a - r_3)}\right) ~ \phi_i.
\end{align}

  \end{widetext}

\section{A posteriori validation of simple poles.}
\label{sec:simple-poles}

We discuss how we convince ourselves that all one-forms in \cref{eq:cDE} have locally only simple poles.
First, we pull back the DE on a random line $\gamma(t)$.
We note that around every singular point $t_i$ one can always choose a suitable constant linear combination of periods
\begin{equation}
  \begin{aligned}
    \hat\psi &= c_1 \psi_1 + c_2 \psi_2, \\
    \hat\phi &= c_1 \phi_1 + c_2 \phi_2, \\
    \hat\pi^{a_i} &= c_1 \pi^{a_i}_1 + c_2 \pi^{a_i}_2,
  \end{aligned}
\end{equation}
such that $\hat\psi$, $\hat\phi$, $\hat\pi^{a_i}$ have converging power series expansions around $t_i$ and in addition the expansion of $\hat\psi$ always starts with a nonzero constant term (see e.g.\ the discussion in \cite{Broedel:2021zij,Duhr:2025lbz}). The latter guarantees that the presence of $\psi$ in denominators cannot introduce any extra poles.
We then observe that every entry of the pulled-back DE has a generic form
\begin{equation} \label{eq:simple-poles}
  \psi^{-k} \sum_i  \psi^{a_i} \phi^{b_i} \pi^{c_i} ~ \dd{\Big(c_i \log{p_i(t)} + q_i(t) \Big)},   
\end{equation}
with constants $c_i$, positive integers $k,a_i,b_i,c_i$, and polynomials $p_i(t), q_i(t)$.
This immediately manifests that only simple poles are possible around any of the singular points along the path $\gamma$.

\bibliography{letter.bib}

\end{document}